\documentclass[arxiv]{iosart2xarxiv}

\usepackage{comment}
\usepackage[dvipsnames]{xcolor}

\definecolor{OliveGreen}{rgb}{0.72, 0.71875, 0.421875}
\definecolor{LightGray}{rgb}{0.97,0.97,0.97}

\usepackage{hyperref}
\usepackage{csquotes}
\hypersetup{
    colorlinks=true,
    linkcolor=blue,
    citecolor=blue,      
    urlcolor=cyan,
    pdftitle={Plasma-KG for LTP Research},
    pdfpagemode=FullScreen,
    }

\usepackage{listings}
\usepackage{inconsolata}
\usepackage{listings} 
\lstdefinelanguage{sparql}{
  alsoletter=-,
  basicstyle=\ttfamily\small,
  backgroundcolor=\color{LightGray},
  columns=fullflexible,
  breaklines=false,
  sensitive=true,
  frame=bt,
  aboveskip=1em,
  belowskip=1em,
  xleftmargin=.5em,
  xrightmargin=.5em,
  framexleftmargin=.5em,
  framextopmargin=.5em,
  framexbottommargin=.5em,
  framexrightmargin=.5em,
  tabsize = 2,
  showstringspaces=false,
  morecomment=[l][\color{gray}]{\#},       
  morecomment=[n][\color{blue}]{<http}{>}, 
  morestring=[b][\color{OliveGreen}]{\'},  
  keywordsprefix=?,
  classoffset=0,
  keywordstyle=\color{Sepia},
  morekeywords={},
  classoffset=1,
  keywordstyle=\color{Purple},
  morekeywords={rdf,rdfs,owl,xsd,purl,plsmo, foaf, vivo},
  classoffset=2,
  keywordstyle=\color{MidnightBlue},
  morekeywords={
    SELECT, DISTINCT,CONSTRUCT,DESCRIBE,ASK,WHERE,FROM,NAMED,PREFIX,BASE,OPTIONAL,
    FILTER,GRAPH,LIMIT,OFFSET,SERVICE,UNION,EXISTS,NOT,BINDINGS,MINUS,a
  }
}

\usepackage{caption}
\DeclareCaptionFont{white}{\color{white}}

\usepackage[most]{tcolorbox}
\definecolor{light-gray}{gray}{0.90}
\tcbset{on line, 
        boxsep=1pt, left=0pt,right=0pt,top=0pt,bottom=0pt,
        colframe=white,colback=light-gray,  
        highlight math style={enhanced}
        }

\pubyear{0000}
\volume{0}
\firstpage{1}
\lastpage{1}

\begin{document}

\begin{frontmatter}

\title{Semantic Information Management in Low-Temperature Plasma Science and Technology with VIVO}
\runtitle{Semantic Information Management in LTP Science and Technology with VIVO}


\begin{aug}
\author{\inits{I.}\fnms{Ihda} \snm{Chaerony Siffa}\ead[label=e1]{ihda.chaeronysiffa@inp-greifswald.de}}
\author{\inits{R.}\fnms{Robert} \snm{Wagner}\ead[label=e2]{second@somewhere.com}}
\author{\inits{M.M.}\fnms{Markus M.} \snm{Becker}\ead[label=e4]{markus.becker@inp-greifswald.de}
\thanks{Corresponding author. \printead{e4}.}}
\address{\orgname{Leibniz Institute for Plasma Science and Technology (INP)}, Greifswald, \cny{Germany}\printead[presep={\\}]{e4}}
\end{aug}

\begin{abstract}
Digital research data management is increasingly integrated across universities and research institutions, addressing the handling of research data throughout its lifecycle according to the FAIR data principles (Findable, Accessible, Interoperable, Reusable). Recent emphasis on the semantic and interlinking aspects of research data, e.g. by using ontologies and knowledge graphs further enhances findability and reusability. This work presents a framework for creating and maintaining a knowledge graph specifically for low-temperature plasma (LTP) science and technology. The framework leverages a domain-specific ontology called Plasma-O, along with the VIVO software as a platform for semantic information management in LTP research. While some research fields are already prepared to use ontologies and knowledge graphs for information management, their application in LTP research is nascent. This work aims to bridge this gap by providing a framework that not only improves research data management but also fosters community participation in building the domain-specific ontology and knowledge graph based on the published materials. The results may also support other research fields in the practical use of knowledge graphs for semantic information management.
\end{abstract}


\end{frontmatter}


\section{Introduction}\label{s1}
The assimilation of digital research data management (RDM) is seeing a growing trend across universities and research institutes. This is visible from the increasing adoption rate of electronic laboratory notebooks (ELNs) and other RDM software tools~\cite{Petersen-2020-ID254, Fink-0000-ID255, Schroder-2022-ID232, ChaeronySiffa-2022-ID147, Jackson-2024-ID303}. This trend focuses not only on the technical aspects of data handling throughout the research lifecycle but also emphasizes the importance of semantic information management in enhancing the findability, understandability, and reusability of research data. RDM practices primarily focus on the technical aspects of data handling, ensuring the availability and accessibility of research data through collection, storage, archiving, and publication on institutional or domain-specific repositories~\cite{Conrad-2024-ID284, Lin-2024-ID283, Rahimzadeh-2023-ID285, Strand-2022-ID302}. However, semantic information management goes beyond organizing and storing data by adding layers of meaning and context through the use ontologies, knowledge graphs, and other semantic technologies~\cite{Schneider-2020-ID256, Schroder-2022-ID232}. This semantic enrichment of research data unlocks its full potential for knowledge discovery and reuse, enabling researchers to identify relationships between datasets, understand the context in which data was collected, and assess its quality and relevance for specific research questions. By making research data machine-readable and understandable, semantic information management enhances the interoperability and reusability of data, aligning with the FAIR (Findable, Accessible, Interoperable, Reusable) data principles~\cite{Wilkinson-2016-ID142}.

\begin{figure}[ht]\centering
\includegraphics[width=1\columnwidth]{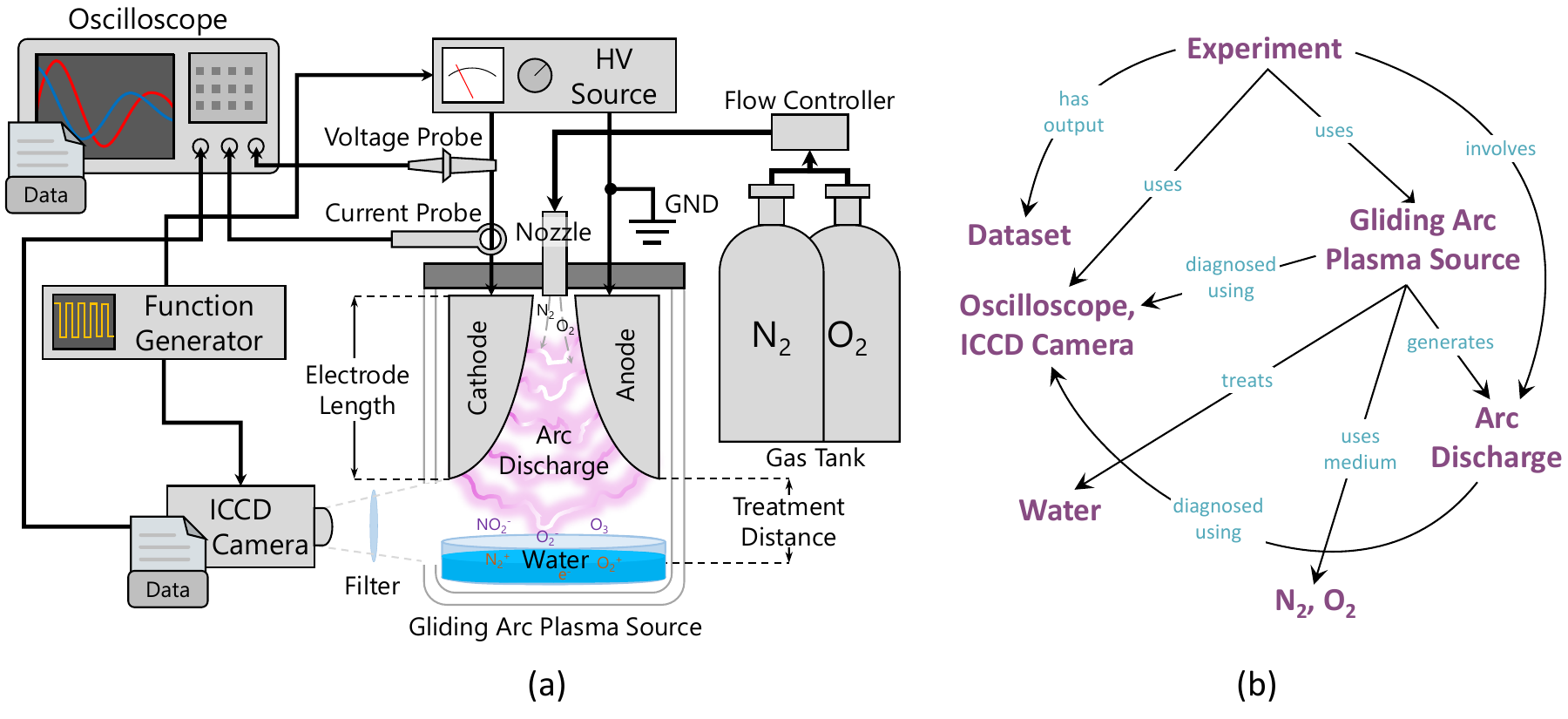} 
\caption{(a) Typical experiment setup found in low-temperature plasma research. (b) Schematic representation of the relationships between the components within the experiment.}
\label{fig:experimentsetup}
\end{figure}

In the field of low-temperature plasma (LTP) science and technology, often characterized by multi-disciplinary research topics and how experiments are conducted (small table-top experiments as opposed to large-scale experiments as common, e.g. in high-energy physics)~\cite{Anirudh-2023-ID171}, the interlinking and explicit relationships between relevant concepts can be practical. For example, a researcher may be interested in finding datasets that are obtained from a certain diagnostic device used to diagnose a certain plasma source. This can be modeled by assigning a relationship between the concepts of plasma source and dataset, i.e. it can be described in an ontology, which is a formal description of concepts or knowledge and their relationships within a domain. In this example, the use of an ontology further improves the findability and comprehensibility of complex data. 
To illustrate further, Figure \ref{fig:experimentsetup}a portrays a setup of a table-top experiment typically used in LTP research, which comprises several interconnected instances (components). Correspondingly, Figure \ref{fig:experimentsetup}b provides a prototypical schematic representation of the relationships between these components, highlighting the interplay between the plasma source, the treated sample, and the diagnostic devices participating in the experiment, which ultimately results in the generation of research data.

The work on ontologies applied in the LTP field is still in its infancy. To the best of our knowledge, there has only been a couple of academic articles on ontology development in the field of computational plasma physics~\cite{Snytnikov-2020-ID119, Sapetina-2020-ID286}, which is adjacent to the LTP field. Notably, the first steps in that direction have been taken within the framework of developing a metadata schema and the foundation of a general knowledge graph for applied plasma physics and plasma medicine~\cite{HaraldProjectReport2023,BeckerProjectReport2023, becker_2020_4091401}.

In contrast, ontologies are already immensely prominent in pharmacology, biomedical sciences, engineering, and enterprises as a way to formalize rules and processes\textemdash provide a common understanding within the corresponding domain~\cite{Woods-2023-ID290, Panzarella-2023-ID263, Calvo-Cidoncha-2022-ID262, Yang-2023-ID261}. Ontologies have also been applied directly in RDM workflows and pipelines. da Silva \textit{et al.}~\cite{RochadaSilva-2014-ID234} have developed the Dendro software platform for preparation of research data, on which the users have the possibility to annotate their data using a given ontology. Schröder \textit{et al.}~\cite{Schroder-2022-ID232} have proposed an RDM workflow utilizing the elabFTW ELN~\cite{CARPi2017} and the PROV ontology~\cite{ProvO2012} to ensure the provenance of the generated research data. Gonçalves \textit{et al.}~\cite{Goncalves-2017-ID233} have developed a web tool driven by an ontology to generate web forms for the acquisition of structured data. It is worth noting that, in the age of Artificial Intelligence (AI), researchers are increasingly highlighting the crucial role of ontologies in enhancing the trustworthiness and accuracy of AI systems~\cite{Allemang-2024-ID257, Fernandez-2023-ID258, Ronzano-2024-ID287}.

This work outlines a framework for constructing and maintaining a knowledge graph in the LTP field. The approach leverages a domain-specific ontology and the VIVO software platform~\cite{Conlon-2019-ID266, Borner-2012-ID236} to represent and manage semantic information. VIVO is an open-source current research information system that utilize ontologies as core data structures. It is popular for managing institutional research information such as researchers, their research outputs and collaboration networks. Websites running on VIVO can be found, e.g. at \url{https://vivo.tib.eu/fis/}, \url{https://vivo.weill.cornell.edu/}, and \url{https://vivo.brown.edu/}. There, VIVO is used as a platform to discover researchers and their expertise, which can be in form of research projects and published papers. Recently, VIVO has been used in managing other kinds of linked open data (LOD) such as historical architectural prints as found in \url{https://sah.tib.eu/}. Additionally, VIVO offers a user-friendly way to ingest data into the knowledge graph, allowing users without technical expertise in graph data to participate in its construction. This fosters a community-maintained, domain-specific knowledge graph.

The rest of the paper is structured as follows. In Section \ref{s2}, we briefly introduce the domain-specific ontology used in this work, the VIVO software platform, and ways to integrate the domain-specific ontology and data instances into VIVO. Subsequently, we present different potential use cases of the suggested research in Section \ref{s3}, specifically in the LTP field. Finally in Section \ref{s4}, we conclude and discuss the outlook of the present work.

\section{Methods}\label{s2}

This section details the development and integration of a domain-specific ontology tailored for the field of low-temperature plasma (LTP) science and technology. The overarching objective is to establish a robust semantic framework that facilitates knowledge creation, discovery, and management within the LTP domain. This is achieved through two primary approaches: the construction of a specialized ontology, Plasma-O, and its subsequent integration into the VIVO platform to create and maintain the knowledge graph.

\subsection{Ontology for Low-Temperature Plasma Science and Technology}\label{s2.1}

The domain-specific ontology presented in this work is built upon the earlier draft~\cite{HaraldProjectReport2023,BeckerProjectReport2023, becker_2020_4091401} and the concepts introduced in the existing plasma metadata schema, named Plasma-MDS, introduced by Franke \textit{et al.}~\cite{Franke-2020-ID168}. There, the introduced metadata schema is intended to support the adoption of the FAIR data principles and RDM in the LTP field, specifically for the publication of datasets generated from LTP experiments. Plasma-MDS is used as the metadata standard for LTP-specific data repositories such as INPTDAT\footnote{\url{https://www.inptdat.de/}}~\cite{Becker-2019-ID260} and RDPCIDAT\footnote{\url{https://rdpcidat.rub.de/}}. The developed ontology in this work, however, is designed to have a broader application range, e.g. as a basis of metadata schemas for various RDM workflows and semantic knowledge creation and discovery. In this work, we focus on the part of the domain-specific ontology for the latter case and consider the first case and the complete ontology for future work. From here on, we shall refer to this domain-specific ontology as the Plasma Ontology or Plasma-O.

While Plasma-MDS serves effectively as a metadata schema for describing and publishing LTP research datasets, Plasma-O, as a formal ontology encompasses a broader scope. Many of the terminologies and concepts present in Plasma-MDS are also defined within Plasma-O, but crucially, Plasma-O extends beyond data description. It establishes formal relationships between these concepts, enabling automated reasoning, inference of new knowledge, and integration of data from various sources. This richer semantic representation allows Plasma-O to support a wider range of RDM workflows, including data analysis, modeling, and complex semantic queries. This helps in facilitating advanced knowledge discovery and turning the FAIR data principles into practice. In essence, Plasma-O is complementary to Plasma-MDS, adding the possibility of semantic linking of data and metadata.

In contrast to Plasma-MDS that introduces \textit{Dataset} as a central class or concept, Plasma-O is built around the central classes \textit{Plasma Study} and \textit{Plasma Experiment}, which are special cases and subclasses of \textit{Scientific Study} and \textit{Scientific Experiment}, respectively. Both may share a similar set of relationships with other classes that makes it natural to group them within the same superclass, i.e. \textit{Research Activity}. Formal expressions can be introduced to ensure the specificity of \textit{Plasma Study} and \textit{Plasma Experiment}, which set them apart from the generic \textit{Scientific Study} and \textit{Scientific Experiment}. For example, the formal expression for \textit{Plasma Experiment} may be written as  
\begin{equation*}
\begin{aligned}[c]
    Plasma\,Experiment &\sqsubseteq \geq 1 involves.Plasma \\
                &\sqsubseteq\exists conductedBy.Agent \\
                &\sqsubseteq\exists hasTopic.Research\,Topic \\
                &\sqsubseteq\exists hasPlasmaSource.Plasma\,Source \\
                &\sqsubseteq\exists hasMedium.Medium \\
                &\sqsubseteq\exists hasTarget.Target \\
                &\sqsubseteq\exists usesMethod.Method \\
                &\sqsubseteq\exists hasOutput.Document \\
                &\sqsubseteq\exists usesDevice.Device \\
\end{aligned}
\end{equation*}
The first property (or relation) of the expression above denotes that every \textit{Plasma Experiment} must involve at least one kind of \textit{Plasma}. The remaining properties relate \textit{Plasma Experiment} to other classes, such as \textit{Plasma Source}, \textit{Medium}, \textit{Target}, \textit{Device}, etc. This is consistent with the concepts introduced in Plasma-MDS~\cite{Franke-2020-ID168}. Referring to Figure \ref{fig:experimentsetup}b, these classes or concepts indeed generalize over various components in the experiment, e.g. the gas $\mathrm{N_2}$ and $\mathrm{O_2}$ can be classified as \textit{Medium}, water as \textit{Target}, gliding arc plasma source as \textit{Plasma Source} (which is modeled as a subclass of \textit{Device}), etc. From the given expression, one can infer the experiment illustrated in Figure \ref{fig:experimentsetup} is of type \textit{Plasma Experiment} considering it involves an arc discharge, which is an instance of \textit{Plasma}. Furthermore, the output of an experiment is \textit{Document}, which may be in form of \textit{Dataset} and/or \textit{Article}, amongst others. Subsequently, one can write an expression of \textit{Plasma Study} as concisely as follows
\begin{equation*}
\begin{aligned}[c]
    Plasma\,Study &\sqsubseteq\left[\left(\geq 1 consistsOf.Plasma\,Experiment\right) \sqcup \left(\geq 1 involves.Plasma \right)\right]
\end{aligned}
\end{equation*}
which means that every \textit{Plasma Study} must have at least one \textit{Plasma Experiment} or involve at least a kind of \textit{Plasma}. Although in practice, this class also shares a similar set of properties like \textit{Plasma Experiment}, which are inherited from the superclass \textit{Research Activity}.

The formal expressions above are presented within the framework of the description logic $\mathcal{ALC}$ \cite{Stephan-2007-ID304} consisting of statements about concepts or classes, instances, and their relations. The symbol $\sqsubseteq$ denotes subsumption, indicating a subclass relationship; for example, $A \sqsubseteq B$ denotes that all instances of concept $A$ are also instances of concept $B$. The notation $\exists r.C$ represents the existential restriction, indicating that there exists at least one relationship $r$ to an instance of concept $C$. So $\exists involves.Plasma$ indicates at that there exists at least an $involves$ relation to an element that belongs to concept/class $Plasma$. The symbol $.$ is part of the description logic syntax to separate properties and fillers. Cardinality restrictions, expressed as $\geq n r.C$, specify the minimum number $n$ of relationships $r$ to instances of concept C. For instance, $\geq 1 consistsOf.PlasmaExperiment$ states that there must be at least one $consistsOf$ relationship to an instance of $PlasmaExperiment$. $\mathcal{ALC}$ also supports conjunction ($\sqcap$, representing intersection, concept AND), disjunction ($\sqcup$, representing union, concept OR), and negation ($\neg$, representing complement, concept NOT).

\begin{figure}[ht]\centering
\includegraphics[width=0.5\columnwidth]{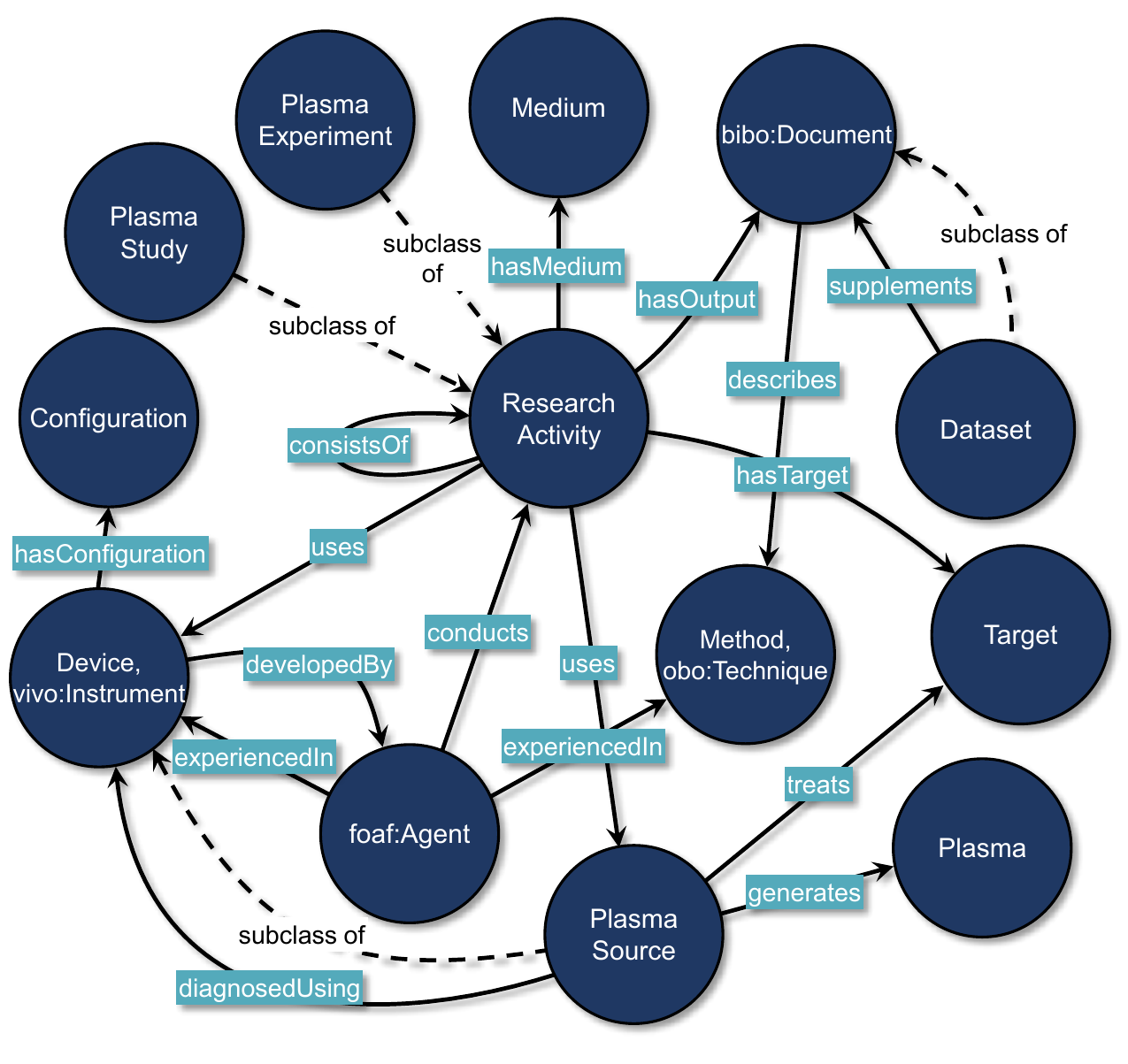} 
\caption{Simplified excerpt of Plasma-O illustrating selected classes and their relationships with each other and classes from external ontologies. Inverse properties are not visualized for simplicity.}
\label{fig:alignment}
\end{figure}

It is worth noting that Plasma-O, by design, includes inverses for most object properties. This means that if \textit{Research Activity} is connected to \textit{Medium} through the \textit{hasMedium} property, the inverse relationship can be expressed using the \textit{mediumOf} property. These inverse properties come in handy when exploring the classes and their instances in Plasma-O and later in the constructed knowledge graph. Figure \ref{fig:alignment} illustrates a simplified representation of Plasma-O consisting of selected classes and their relationships relevant for the manuscript. Furthermore, Table \ref{table:class-descriptions} provides descriptions of the classes. 

\begin{table}[ht]
\caption{\label{table:class-descriptions}Descriptions of selected classes from Plasma-O and reused classes from other ontologies.}
\centering
\begin{tabular}{ l l p{10cm}  }
\hline
 Class & Ontology & Description \\
\hline
 \textit{Research Activity} & Plasma-O & Any systematic investigation or inquiry aimed at increasing knowledge or understanding. \\
 \textit{Plasma Experiment} & Plasma-O & A scientific experiment that involves plasma as the object of investigation and/or its use in various kinds of applications, e.g. plasma medicine, decontamination, material processing, etc. \\
 \textit{Plasma Study} & Plasma-O & A collection of experiments with the focus on investigating plasma behaviors/properties, effects of a plasma on a certain target, its applications, etc. \\
 \textit{Device} & Plasma-O & A device is an object that is invented or constructed to serve a specific function. Also equivalent to vivo:\textit{Instrument}. \\
 \textit{Plasma Source} & Plasma-O & A device that generates a plasma. It is a subclass of \textit{Device} \\
 \textit{Medium} & Plasma-O & A medium such as gas, mixture of gases, or other substance, which is either (partially) ionized or otherwise activated by a plasma source. \\
 \textit{Target} & Plasma-O & An entity, which a plasma acts on. \\
 \textit{Method} & Plasma-O & Equivalent to obo:\textit{Technique}: a technique is a planned process used to accomplish a specific activity or task. \\
 \textit{Configuration} & Plasma-O & A set of settings and parameters of a device or instrument. \\
 \textit{Agent} & FOAF & Agents are things that do stuff. Subclasses of this class include \textit{Person}, \textit{Institute}, \textit{Group}, etc. \\
 \textit{Document} & BIBO & A document (noun) is a bounded physical representation of body of information designed with the capacity (and usually intent) to communicate. A document may manifest symbolic, diagrammatic or sensory-representational information. \\
 \textit{Dataset} & VIVO & Dataset is a document consisting of a collection of digital resources obtained from scientific experiments. It is a subclass of \textit{Dataset}. \\
\hline
\end{tabular}
\end{table}

Plasma-O can be attached to other ontologies by means of reusing existing classes from them. Here, we attach Plasma-O to the VIVO ontology~\cite{Corson-Rikert-2012-ID235}, which forms the backbone of the VIVO platform. The VIVO ontology itself reuses classes and properties from other ontologies such as FOAF~\cite{Brickley2014}, BIBO~\cite{DArcus2009}, and OBO~\cite{Smith-2007-ID265}, to name a few. Together, these ontologies are used to describe various different concepts that can be grouped into people, activities, courses, events, organizations, equipment, research, and locations. This encompasses the general concepts in most scientific disciplines and highlights the interoperability of the different concepts from these ontologies. Some concrete examples include the reuse of \textit{Agent} from FOAF connected to \textit{Plasma Experiment} via the \textit{conductedBy} property, and \textit{Document} from BIBO via \textit{hasOutput}. Apart from reuse, Plasma-O also considers class equivalences, e.g. \textit{Method} is equivalent to \textit{Technique} from OBO, and \textit{Device} is equivalent to \textit{Instrument} from VIVO.

Despite being able to describe general relevant concepts in LTP research, Plasma-O is still being developed further. The presented work strives to provide public access to the ontology and related knowledge graph going---with respect to accessibility for humans and machines---far beyond storing the source files in a public repository. The suggested approach of using VIVO as a semantic knowledge management system for domain-specific data and metadata (beyond its original application scenario) is seen as transferable to other domains. With this, it potentially fosters the adoption of the FAIR data principles in different interdisciplinary research communities. The version of Plasma-O introduced in this work (version 0.7.0)\footnote{Release version 0.7.0 is available at \url{https://github.com/plasma-mds/plasma-ontology/releases/tag/v0.7.0}} consists of 61 classes, 105 object properties, and 12 individuals whose numbers are expected to grow in the next iteration of the ontology.

\subsection{LTP Knowledge Graph Development with VIVO}\label{s2.2}
The core technology of VIVO is the ontologies. VIVO provides a user-friendly user interface (UI) that enables end users (e.g. data curators) to create and manage semantic information in the form of a knowledge graph based on the available ontologies. As mentioned in Section \ref{s2.1}, VIVO comes with the VIVO ontology and various other ontologies readily facilitating the creation of a knowledge graph representing researchers, their interests, projects, and outputs. Additionally, incorporating new ontologies and/or knowledge graphs into VIVO is simplified through its add/remove RDF data feature. Given the aforementioned features and existing resources in VIVO, we see the opportunity to extend VIVO's original use case that is current research information system with a more detailed semantic knowledge management system for research objects in LTP research. Compared to other systems such as Semantic MediaWiki\footnote{\url{https://www.semantic-mediawiki.org}} and Wikibase\footnote{\url{https://wikiba.se/}}, VIVO offers out-of-the-box flexibility and simplicity for content integration and reuse for both humans and machines, which we see as a big advantage. 

Accordingly, we introduce Plasma-O into VIVO and set the namespace (or prefix) to \enquote{plsmo} for quick identification of the ownership of the classes. Note that, we use VIVO version 1.15 for this work. For a more detailed documentation to install and set up VIVO version 1.15, we refer the readers to the relevant VIVO documentation\footnote{\url{https://wiki.lyrasis.org/display/VIVODOC115x}}.

\begin{figure}[ht]\centering
\includegraphics[width=1\columnwidth]{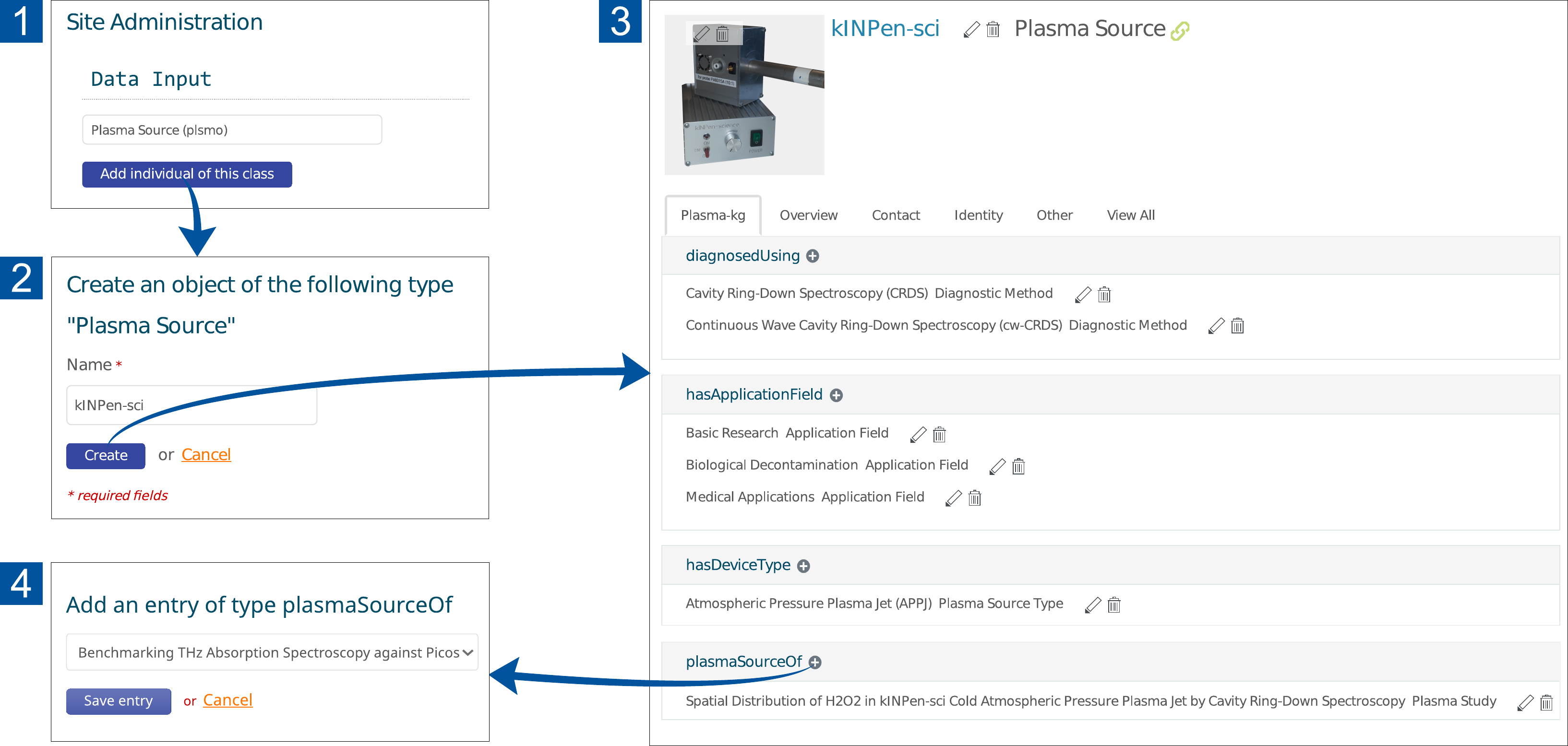} 
\caption{Example workflow for creating an instance of \textit{Plasma Source}.}
\label{fig:create-source}
\end{figure}

There are several ways to populate the ontologies with instances or data on VIVO to create a knowledge graph \textit{ab intio}. The most straightforward way is to use the data input feature. As a first step, we create a class group, called \enquote{plsmo}, and pick relevant Plasma-O classes (and also relevant classes from other related ontologies) to be included in the plsmo group. This is done to ease the data ingest process on VIVO. We can then choose the starting class we want to populate with instances. Figure \ref{fig:create-source} shows an example workflow for creating an instance of \textit{Plasma Source}. It starts from selecting the \textit{Plasma Source} class (available under the plasma-o group). Then we will be prompted to provide a name for this instance, in this case, we choose the technical name for this plasma source, which is \enquote{kINPen-sci}. Upon creating the instance, we will be directed to the newly created kINPen-sci instance page as shown in step 3 in Figure \ref{fig:create-source}. Since \textit{Plasma Source} is connected to different other classes in the ontology through the corresponding properties, these connections will be automatically displayed on the kINPen-sci instance page. Here, we can proceed to link the kINPen-sci instance with either existing instances or new instances. In the example, we link the kINPen-sci instance with an already existing \textit{Plasma Study} instance, through the \textit{plasmaSourceOf} property. A similar process can be done for other classes and properties. In addition to creating the instance one by one, which can be time consuming, VIVO also allows for creation of instances and linking in bulk using CSV files. This particular approach streamlines the process by allowing users to provide instance details like id, name, uri, and other relevant properties in a structured format.

Regardless of the creation method, utilizing ontologies within VIVO maximizes the reuse of existing instances to describe various concepts. This approach is applicable to all classes across all ontologies available in VIVO, and the created instances inherit the relationships defined within those ontologies. These interconnected instances form the foundation of the knowledge graph. Here, we constructed the plasma knowledge graph (Plasma-KG) by applying these workflows and incorporating data from the INPTDAT data repository\footnote{\url{https://www.inptdat.de}}~\cite{Becker-2019-ID260} and various publications in the LTP field.

\section{Results}\label{s3}

The primary and most general use case of VIVO is to create profiles for researchers and their research outputs, typically in the form of published academic articles and other contributions. This directly relates to the expertise these researchers possess. For instance, a researcher might be linked to several publications and listed as experienced in specific plasma sources, devices, or diagnostic methods. Beyond this, we demonstrate other potential use cases of VIVO within the LTP field, which can be adapted by other domains as well.

\subsection{Semantic Cataloging and Browsing of Domain-Specific Information and Data}\label{s3.1}
This use case extends the general use case of VIVO that is researcher profiling. Using Plasma-O, we can create a semantic catalog for research objects relevant in the LTP field such as plasma sources and datasets generated within the framework of plasma studies. This use case attempts to semantically enrich the existing solution offered by the INPTDAT data repository~\cite{Becker-2019-ID260}, where researchers can share datasets from their experiments or studies as well as the details of the plasma sources they use or develop or related applications and patents.

\begin{figure*}[htbp]\centering
\includegraphics[width=1\textwidth]{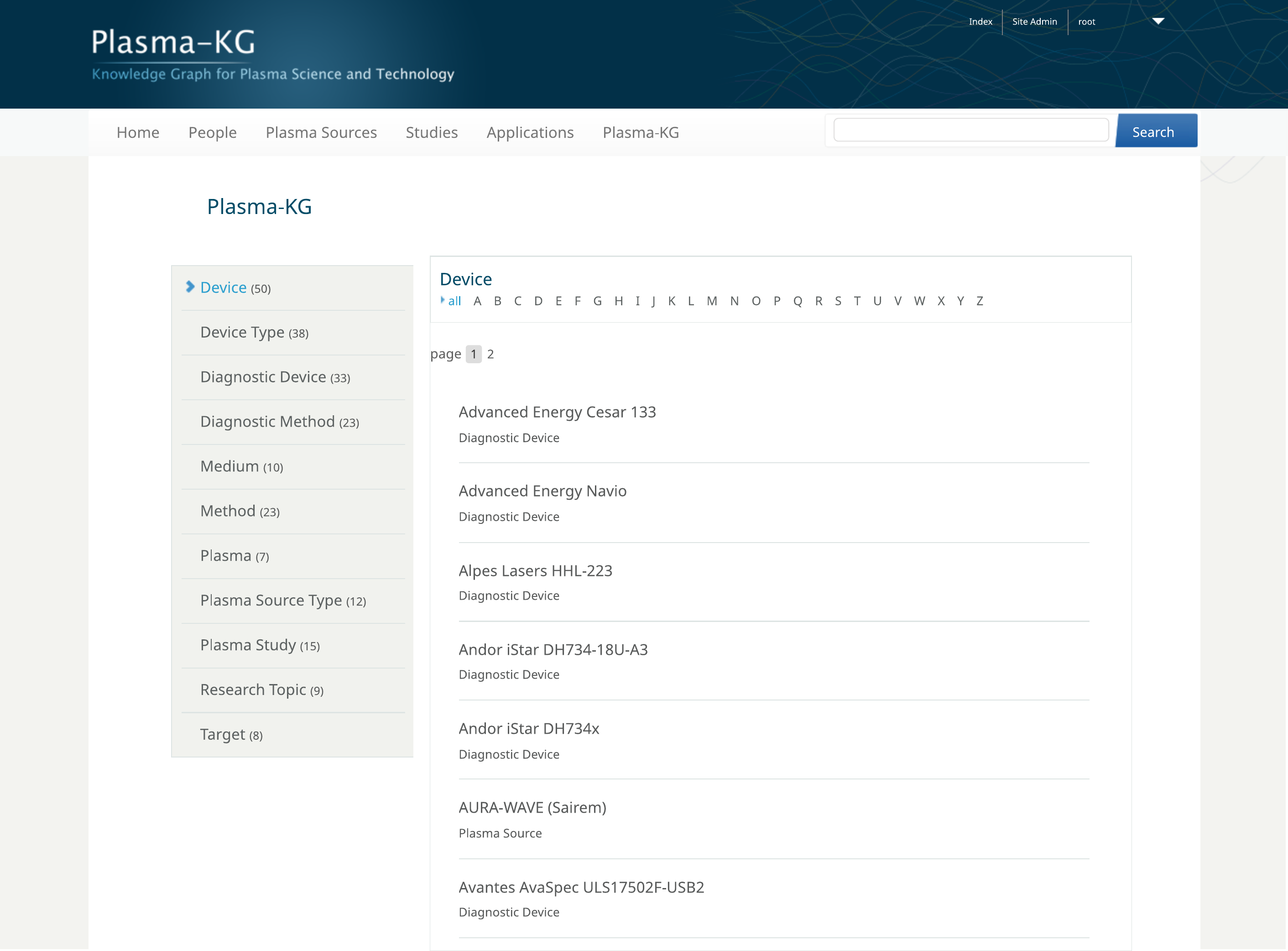} 
\caption{Example of a device catalog within the plasma knowledge graph showcasing plasma sources and diagnostic devices. Faceted browsing feature enhances navigation across the semantic information landscape of the LTP domain.}
\label{fig:navigability}
\end{figure*}

INPTDAT is a DKAN\footnote{\url{https://demo.getdkan.org/}}-based data portal utilizing the relational database technology which does not directly support relationships between data points. In contrast, ontology-based software like VIVO automatically interlinks instances of classes like \textit{Plasma Source} and \textit{Dataset} based on the applied ontology, which enriches the connections between these classes and other relevant classes. This results in enhanced navigability while browsing these research objects, which is highly relevant in terms of the FAIR data principles. Figure \ref{fig:navigability} showcases an example device catalog within VIVO along with its faceted browsing feature, where researchers can browse devices used in LTP research, including plasma sources and diagnostic devices. Additionally, researchers can navigate through various lists of instances based on their class e.g. plasma studies, targets, media, etc.

\subsection{Semantic Information Retrieval Using SPARQL}\label{s3.2}
VIVO offers a user friendly way to ingest data into the knowledge graph. This allows users that are not experienced in the technicalities of graph data to participate in building the knowledge graph, which results in a community-maintained (domain specific) knowledge graph. Once the knowledge graph contains enough data, we can construct complex competency questions that can be implemented using SPARQL\footnote{\url{https://www.w3.org/TR/sparql11-query/}}. This use case is not specific to VIVO, nonetheless, VIVO offers a SPARQL query API that allows a self-contained solution for an ontology-based information management system.

While ingesting data using VIVO is straightforward for general users, formulating complex competency questions using SPARQL requires some knowledge in the query language and the ontology itself. Automated processes to represents the questions in a natural language into SPARQL queries\footnote{\url{https://github.com/LiberAI/NSpM}}~\cite{Yin-2021-ID264}, by which non-technical users can understand, are a topic on its own and beyond the scope of the present work. Nevertheless, we demonstrate this use case in this work by constructing two competency questions that are relevant in the LTP field as follows:

\begin{enumerate}
    \item \enquote{Who is experienced in \textit{Plasma Source} at \textit{Institute}?}
    \item \enquote{Which \textit{Dataset} is obtained with \textit{Method} applied to diagnose \textit{Plasma Source}?}
\end{enumerate}

The SPARQL query for the first competency question is shown in Listing \ref{sparql-query-1}. This query selects \textit{Plasma Source} and \textit{Institute} instances with the labels \enquote{kINPen-sci} and \enquote{INP Greifswald}, respectively. It then infers instances of foaf:\textit{Person} (a subclass of foaf:\textit{Agent}) related to the identified \textit{Plasma Source} and \textit{Institute}. The query results, presented in Table \ref{table:sparql-result-1}, reveal four foaf:\textit{Person} instances (with their ids and labels) that have experience with \enquote{kINPen-sci} (instance id: n6807) and affiliated with \enquote{INP Greifswald} (instance id: n3601).

\begin{lstlisting}[language=sparql, escapechar=^, label=sparql-query-1,caption=SPARQL query for \enquote{Who is experienced in \textit{Plasma Source} at \textit{Institute}?}]
PREFIX rdfs: <http://www.w3.org/2000/01/rdf-schema#>
PREFIX rdf:   <http://www.w3.org/1999/02/22-rdf-syntax-ns#>
PREFIX plsmo: <https://plasma-mds.org/ontology/plasma-o/>
PREFIX foaf: <http://xmlns.com/foaf/0.1/>
PREFIX vivo: <http://vivoweb.org/ontology/core#>
#
# Who is experienced in the plasma source "kINPen-sci"
# at the institute "INP Greifswald?"
#

SELECT DISTINCT ?Person ?PersonLabel ?PlasmaSource ?Institute 

WHERE
{
  # Define the types of the involved entities
  ?Person rdf:type foaf:Person .
  ?Institute rdf:type vivo:Institute .
  ?PlasmaSource rdf:type plsmo:PLSMO_C0005 .
  
  # Get the instance with label "INP Greifswald"
  ?Institute rdfs:label "INP Greifswald"@en-US .
  # Get the instance with label "kINPen-sci"
  ?PlasmaSource rdfs:label "kINPen-sci"@en-US .
  
  # Main query
  
  # Get Person and Institute instances that are connected with
  # a predicate "affiliatedWith" which has IRI plsmo:PLSMO_R0003
  ?Person plsmo:PLSMO_R0003 ?Institute .
  
  # Get Person and PlasmaSource instances that are connected with
  # a predicate "experiencedIn" which has IRI plsmo:PLSMO_R0032
  ?Person plsmo:PLSMO_R0032 ?PlasmaSource .
  
  # Showing the labels for Person
  ?Person rdfs:label ?PersonLabel .  									
}
# Returns maximum of 5 results
LIMIT 5
\end{lstlisting}

\label{appendix:sparql-results}
\begin{table}[ht]
\caption{\label{table:sparql-result-1}SPARQL results for \enquote{Who is experienced in \textit{Plasma Source} at \textit{Institute}?} (Listing \ref{sparql-query-1}). Note that the names are redacted for privacy reasons.}
\centering
\begin{tabular}{ l l l l l  }
\hline
 Person & PersonLabel & PlasmaSource & Institute \\
\hline
 n1951 & N***, A*** *. *. & n6807 & n3601 \\
 n3571 & G******, T****** & n6807 & n3601  \\
 n4703 & K****, S****-J****** & n6807 & n3601\\
 n2146 & K***, L**** &   n6807 & n3601 \\
\hline
\end{tabular}
\end{table}
Similarly, Listing \ref{sparql-query-2} represents the SPARQL query for the second competency question. In this case, the query selects \textit{Method} and \textit{Plasma Source} instances with the labels \enquote{Cavity Ring-Down Spectroscopy} and \enquote{kINPen-sci}, respectively, to infer the related datasets (of class \textit{Dataset}). The \textit{Dataset} class is connected with \textit{Method} and \textit{Plasma Source} classes mainly through \textit{Plasma Study}. Therefore, the query first infers the plasma studies that are related to the selected \textit{Method} and \textit{Plasma Source} using the relevant properties, then finally, it infers the datasets related to the inferred plasma studies. Similarly, the answer of this query is shown in Table \ref{table:sparql-result-2}, which returns two datasets.
\begin{lstlisting}[language=sparql, escapechar=^, label=sparql-query-2,caption=SPARQL query for \enquote{Which \textit{Dataset} is obtained with \textit{Method} applied to diagnose \textit{Plasma Source}?}]
PREFIX rdfs:        <http://www.w3.org/2000/01/rdf-schema#>
PREFIX rdf:   <http://www.w3.org/1999/02/22-rdf-syntax-ns#>
PREFIX plsmo:  <https://plasma-mds.org/ontology/plasma-o/>
PREFIX vivo: <http://vivoweb.org/ontology/core#>

#
# Which datasets are obtained with the diagnostic 
# method "Cavity Ring-Down Spectroscopy applied (CRDS)" 
# to diagnose the plasma source "kINPen-sci?"
#

SELECT DISTINCT ?Dataset ?DatasetLabel ?PlasmaStudy ?Method ?PlasmaSource

WHERE
{
  # Define the types of the involved entities
  ?Dataset rdf:type vivo:Dataset .
  ?PlasmaStudy rdf:type plsmo:PLSMO_C0024 .
  ?Method rdf:type plsmo:PLSMO_C0012 .
  ?PlasmaSource rdf:type plsmo:PLSMO_C0005 .

  # Get the instance with label "Cavity Ring-Down Spectroscopy (CRDS)"
  ?Method rdfs:label "Cavity Ring-Down Spectroscopy (CRDS)"@en-US .
  # Get the instance with label "kINPen-sci"
  ?PlasmaSource rdfs:label "kINPen-sci"@en-US .
  
  # Main query
  
  # Get instances of PlasmaStudy and Method connected
  # via a predicate "usesMethod" which has IRI plsmo:PLSMO_R0105
  ?PlasmaStudy plsmo:PLSMO_R0105 ?Method .
  
  # Get instances of PlasmaStudy and PlasmaSource connected
  # via a predicate "hasPlasmaSource" which has IRI plsmo:PLSMO_R0039
  ?PlasmaStudy plsmo:PLSMO_R0039 ?PlasmaSource .
  
  # Finally, get instances of Dataset and PlasmaStudy connected
  # via a predicate "outputOf" which has IRI plsmo:PLSMO_R0062
  ?Dataset plsmo:PLSMO_R0062 ?PlasmaStudy .
  
  # Note: Dataset is connected to PlasmaStudy by the outputOf property,
  # which is the inverse of the hasOutput property
  
  # Showing the labels for Dataset
  ?Dataset rdfs:label ?DatasetLabel .  				
}
# Returns maximum of 5 results
LIMIT 5
\end{lstlisting}

\begin{table}[ht]
\caption{\label{table:sparql-result-2}SPARQL results for \enquote{Which \textit{Dataset} is obtained with \textit{Method} applied to diagnose \textit{Plasma Source}?} (Listing \ref{sparql-query-2}).}
\centering
\begin{tabular}{ l l l l l l  }
\hline
 Dataset & DatasetLabel & PlasmaStudy & Method & PlasmaSource \\
\hline
 n5263 & The spatial d... & n4109 & n6324 & n6807\\
 n1597 & The localised... & n1186 & n6324 & n6807\\
\hline
\end{tabular}
\end{table}

\section{Conclusion and Outlook} \label{s4}
The present work introduces the initial foundation of a domain-specific ontology for low-temperature plasma science and technology, named Plasma-O, and its application using the VIVO platform to build a knowledge graph for the LTP field, named Plasma-KG. By aligning Plasma-O with the existing VIVO ontology, we expand VIVO's capabilities to include specialized applications like the semantic cataloging of plasma sources and datasets, and the ability to address complex research questions through SPARQL queries. These enhanced capabilities directly contribute to the adoption of the FAIR data principles in the LTP field  and connects the often disjunct topics of current research information management, research data management and semantic knowledge management in an interoperable way. The improved discoverability and findability of complex data facilitated by Plasma-O and VIVO significantly enhance the Findability and Accessibility aspects of FAIR data principles. Moreover, the standardized, machine-readable format of ontologies and knowledge graphs inherently promotes Interoperability and Reusability of LTP knowledge and research data. The presented methods provide a generalizable platform for domain-specific semantics, which can also be used by third-party services, e.g. to annotate research assets such as patents and scientific literature~\cite{Aras-2024-ID288}. Furthermore, the methods are readily adaptable to other adjacent applied research fields, in which domain-specific ontologies may have been developed. The potential for alignment with Plasma-O exists where relevant. For example, Plasma-O provides the basis to semantically integrate more generic metadata schemas like Plasma-MDS with more specific schemas, e.g. OpenPMD for particle simulations \cite{huebl_2017_822396}. Moreover, Plasma-O allows to connect the concepts from LTP research with adjacent domains, e.g. material science via PMDco \cite{Bayerlein-2024-ID305}, and particle physics \cite{Konnecke-2015-ID306}. Most importantly, this work aims to stimulate related discussions within the LTP community, hopefully steering the future development of the ontology and knowledge graph.

Currently, Plasma-O models general concepts in the LTP field, which limits the types and complexity of competency questions that can be formulated. In future work, we intend to expand the ontology to cover fine-grained details of plasma studies and experiments (see Fig. \ref{fig:experimentsetup}), such as \textit{Property} of \textit{Target} and \textit{Plasma}, \textit{Reactive Species} (e.g. $\mathrm{N_2^+}$ and $\mathrm{O_2^-}$) produced from the interaction between \textit{Plasma} and \textit{Target} (e.g. water), physical quantities such as \textit{ElectrodeLength} and \textit{TreatmentDistance}, etc. This could be relevant not only for knowledge discovery but also during the experimentation stage of research. Consequently, we plan to integrate the ontology and knowledge graph into an RDM tool like Adamant\footnote{\url{https://github.com/plasma-mds/adamant}}~\cite{ChaeronySiffa-2022-ID147}. This would allow researchers to interact with the ontology, thus the knowledge graph, as early as the experimentation stage (e.g. while documenting the experiments in the electronic laboratory notebook), enabling them to receive suggestions or reuse existing metadata for their experiments.

\section*{Code and Data Availability}
\label{sec:data_statement}
The VIVO software platform version 1.15 is used in this work and publicly available at the corresponding GitHub release page\footnote{\url{https://github.com/vivo-project/VIVO/releases/tag/vivo-1.15.0}}. The Plasma Ontology and the example SPARQL queries presented in this work are publicly available at the corresponding GitHub page\footnote{\url{https://github.com/plasma-mds/plasma-ontology}}. The plasma ontology (version 0.7.0) and knowledge graph used in this work are available at: \url{https://doi.org/10.5281/zenodo.14610064}.
The VIVO platform containing the plasma knowledge graph is openly available at: \url{https://vivo.plasma-mds.org}.

\section*{Acknowledgment}
\label{sec:acknowledgement}
This work was partly funded by the Deutsche Forschungsgemeinschaft (DFG, German Research Foundation)---Project Number 496963457. 





\nocite{*}
\bibliographystyle{ios2-nameyear}  
\bibliography{main}        

%

\end{document}